\documentclass[twocolumn,amsmath,amssymb,]{revtex4}
\usepackage{graphicx}% Include figure files
\usepackage{dcolumn}% Align table columns on decimal point
\usepackage{bm}% bold math
\newcommand{\f}{\begin{equation}}
\newcommand{\ff}{\end{equation}}
\newcommand{\fa}{\begin{eqnarray}}
\newcommand{\ffa}{\end{eqnarray}}
\begin{document}

\preprint{}

\title{The Big Bounce in Rainbow Universe}
\author{Yi Ling${}^{1,2}$}
 \email{yling@ncu.edu.cn}
\author{Qingzhang Wu ${}^{1}$}
 \email{ytsywqzz@163.com}
 \affiliation{${}^{1}$ Center for Relativistic Astrophysics and High Energy
Physics, Department of Physics, Nanchang University, Nanchang
330031, China } \affiliation{${}^{2}$ Kavli Institute for
Theoretical Physics China(KITPC), Beijing 100080, China}

\begin{abstract}
The usual Einstein's equations is modified as a one parameter
family of equations in the framework of rainbow gravity. In this
paper we derive the modified Friedmann-Robertson-Walker (FRW)
equations when the cosmological evolution of radiation particles
is taken into account. In particular, given some specific
dispersion relations, the big bounce solutions to the modified FRW
equations can be derived. Notably, to obtain a well defined
rainbow metric at the moment of the big bounce, we find it seems
necessary to introduce a cosmological constant which depends on
the energy of probes as well, implying that a universe with a
positive cosmological constant more likely undergoes a big bounce
at least at this phenomenological level.

\end{abstract}

\pacs{Valid PACS appear here} %PACS, %the Physics and Astronomy
%Classification Scheme.
%Use showkeys class option if keyword
                              %display desired
\maketitle

\section{Introduction}
Before a complete and fundamental quantum theory of gravity can be
established, it has been suggested that a semi-classical or
phenomenological theory of quantum gravity may play a crucial role
in disclosing the nature of quantum gravity effects, in particular
its possible impact on the very early universe and extremely high
energy physics
\cite{Amelino00ge,Amelino00mn,Amelino03ex,Amelino03uc,Magueijo01cr,
Magueijo02am,Smolin02sz,Smolin05cz,Ghosh06,AmelinoCamelia05ik}.
Recently such a semi-classical formalism named as rainbow gravity
has been proposed by Magueijo and Smolin, which can also be viewed
as an extension of doubly special relativity in curved
spacetime\cite{Magueijo02xx}. In this formalism, one key
ingredient is that there is no single fixed spacetime background
when the quantum gravity effects of moving probes on geometry is
taken into account. Instead, it is replaced by a one parameter
family of metrics which depends on the energy of these probes,
forming a ``rainbow'' metric. More explicitly, suppose the
modified dispersion relation in doubly special relativity has a
general form
\begin{equation}
\varepsilon^2f^2(l_p\varepsilon)-p^2g^2(l_p\varepsilon)=m^2_0,
\end{equation}
where two general rainbow functions $f^2(\varepsilon)$ and
$g^2(\varepsilon)$ depend on the energy $\varepsilon$ of probes
and may be expanded in the order of Planck length $l_p= \sqrt{8
\pi G}\sim 1/M_p$. Obviously one requires that
$f^2(\varepsilon),g^2(\varepsilon)\sim 1$ as $\varepsilon/ M_p\ll
1$ . Correspondingly, it is conjectured that the usual flat metric
should be replaced by the rainbow metric defined as
\begin{equation}
ds^2=-\frac{1}{f^2(\varepsilon)}dt^2+\frac{1}{g^2(\varepsilon)}dx^2,\label{frm}
\end{equation}
such that the contraction between infinitesimal displacement and
momenta is a linear invariant\cite{Kimberly03hp,Magueijo02xx}.
\begin{equation}
dx^\mu{p_\mu}=dt\varepsilon+dx^jp_j.
\end{equation}

The flat rainbow metric (\ref{frm}) indicates that the geometry of
spacetime depends on the energy of a particle moving on it. That
is to say, even in the absence of gravity a moving particle with
energy $\varepsilon$ would probe a geometry which is described by
an energy-dependent set of orthonormal frame fields
\begin{equation}
e_0=f^{-1}(l_p\varepsilon)\tilde{e}_0, \ \ \ \ e_i
=g^{-1}(l_p\varepsilon)\tilde{e}_i,
\end{equation}
where the tilde quantities are the energy independent frame fields
probed by low energy particles. In this manner the metric can be
written as
\begin{equation}
g(\varepsilon)=\eta^{ab}e_a(\varepsilon)\otimes e_b(\varepsilon),
\end{equation}
which can be viewed as a one family of flat metrics, characterized
by $\varepsilon$. This strategy overcomes the difficulty of
defining the position space conjugate to the momentum space
arising in $DSR$ where the Lorentz symmetry is accomplished by
nonlinear transformations in momentum space(For recent progress
and discussion on this issue we refer to \cite{Hossenfelder06rr}).

Rainbow metric formalism can be pushed forward when the gravity is
taken into account. First of all, a deformed equivalence principle
of general relativity is proposed in \cite{Magueijo02xx},
requiring that the free falling observers who make measurements
with energy $\varepsilon$ will observe the same laws of physics as
in doubly special relativity. As a consequence, the background
spacetime is described by a general rainbow metric
$g_{\mu\nu}(\varepsilon)$. Then, through the standard process the
corresponding one-parameter family of connection
$\nabla(\varepsilon)_{\mu}$ which is compatible with the rainbow
metric $g_{\mu\nu}(\varepsilon)$ and the curvature tensor
$R(\varepsilon)^\rho_{\mu\nu\lambda}$ can be constructed, leading
to a family of modified Einstein's equations
\begin{equation}
G_\mu{_\nu}(\varepsilon)=8\pi{G(\varepsilon)}T_\mu{_\nu}(\varepsilon)
+g_\mu{_\nu}(\varepsilon)\Lambda(\varepsilon)\label{eq4},
\end{equation}
where the Newton's constant $G(\varepsilon)$ and the cosmological
constant $\Lambda(\varepsilon)$ are expected to vary with the
energy $\varepsilon$ as well from a renormalization group point of
view.

Rainbow gravity formalism has received a lot of attention recently
and other stimulating work on this formalism can be found, for
instance, in
\cite{Galan04st,Galan05ju,Hackett05mb,Aloisio05qt,Ling05bp,Galan06by,Ling06ba,Ling06az,Girelli06fw,Liu07fk,Peng07nj,Li08gs}.
In particular, in \cite{Ling06az} one of the authors generalized
the modified Friedmann-Robertson-Walker(FRW) equations originally
presented in \cite{Magueijo02xx} by considering the cosmological
evolution of probes and derived solutions in which the spacetime
curvature has an upper bound such that the cosmological
singularity is absent. But in \cite{Ling06az} the big bounce
solution to the modified $FRW$ equations is not available, which
greatly depends on what kind of modified dispersion relations we
would apply.

Another motivation of looking for non-singular bouncing solutions
in rainbow gravity formalism comes from recent progress in
cosmology. It suggests that bouncing universes could play a
similar role as inflationary scenario in solving the well known
problems in standard Big-Bang cosmology. Nowadays there are a lot
of cosmological models with a bounce solution such as the pre
Big-Bang scenario \cite{Gasperini02bn}and cyclic/Ekpyrotic
universe \cite{Khoury01wf,Steinhardt01st}, as well as superstring
cosmology\cite{Lidsey99mc,Constantinidis99cu,Tsujikawa03pn,Brandenberger98zs,Fabris02pm},
brane cosmology\cite{Dvali00hr,Shtanov02mb}, loop quantum
cosmology \cite{Bojowald01xe,Stachowiak06uh,Mielczarek08zv} and
quintom models \cite{Feng04ad,Cai07qw}. For more details on
bouncing universe we refer to a recent review\cite{Novello08ra}
and references therein. Since rainbow gravity formalism is
proposed as a semi-classical theory in which the quantum effects
of gravity is taken into account, we would like to ask if it is
possible to obtain such big bounce solutions in this framework.
Bearing this question in mind, in this paper we intend to further
investigate the rainbow universe in a more general setting.
Through some explicit model constructions we find the answer is
affirmative.

We organize the paper as follows. In section II, we derive the
modified $FRW$ equations in the framework of rainbow universe
where the cosmological evolution of probes is taken into account.
Then we turn to derive the big bounce solutions to these equations
with vanishing cosmological constant in section III. Through
specifying modified dispersion relations, two sorts of big bounce
solutions are demonstrated. But these models contain some
unsatisfactory features at the moment when the universe passing
through the big bounce. In section IV we show that such
unsatisfactory points can be overcome by introducing an effective
cosmological constant. The corresponding big bounce solutions are
presented as well.

\section{Modified FRW Universe in rainbow gravity formalism}
The modified $FRW$ equations in rainbow gravity formalism have
originally been derived in \cite{Magueijo02xx}. As a starting
point, the conventional FRW metric is replaced by a rainbow metric
parameterized by the energy $\varepsilon$ of probes
\begin{equation}
ds^2=-\frac{1}{f^2(\varepsilon)}dt^2+\frac{a^2}{g^2(\varepsilon)}\delta_i{_j}dx^idx^j\label{eq5}.
\end{equation}
Here we only consider the spatially flat case with $K=0$. This
metric is defined in a general sense in that one is free to pick up
arbitrary particle as a probe and for any specific measurement its
energy $\varepsilon$ can be treated as a constant which is
independent of spacetime coordinates. However, in \cite{Ling06az}
rather than considering any specific measurement, this mechanism is
generalized to study the semi-classical effects of particles on the
background metric during a longtime process. Then the probe energy
appearing in the rainbow metric is identified with one of photons or
other sorts of massless particles like gravitons or inflatons which
dominate the very early universe. Thus the evolution of energy
$\varepsilon$ with the cosmological time need to be taken into
account, denoted as $\varepsilon(t)$. As a result the rainbow
functions $f$ and $g$ in Eq.(\ref{eq5}) depend on time only
implicitly through the energy of particles. In \cite{Ling06az} we
take the ansatz with $g^2=1$. Here for our purpose $f$ and $g$ are
chosen a priori and we will derive the modified $FRW$ equations in a
more general setting.

Directly starting from the rainbow metric in Eq.(\ref{eq5}) and
taking the cosmological evolution of $\varepsilon$ into account, we
obtain the non-vanishing components of associated connection as
\begin{eqnarray}
\Gamma^0{_0{_0}}&=&-\frac{\dot{f}}{f},
\Gamma^i{_j{_0}}=\left(\frac{\dot{a}}{a}-\frac{\dot{g}}{g}\right)\delta_i{_j},\nonumber\\
\Gamma^0{_i{_j}}&=&\frac{f^2a^2}{g^2}\left(\frac{\dot{a}}{a}-\frac{\dot{g}}{g}\right)\delta_i{_j}.
\end{eqnarray}
Next the components of Ricci tensor as well as the Ricci scalar
can be derived as follows
\begin{eqnarray}
R_0{_0}&=&-3\left(\frac{\ddot{a}}{a}-\frac{\ddot{g}}{g}+\frac{\dot{a}\dot{f}}{af}-\frac{\dot{f}\dot{g}}{fg}-\frac{2\dot{a}\dot{g}}{ag}+\frac{2\dot{g}^2}{g^2}\right),\\
R_i{_j}&=&\frac{f\dot{f}a\dot{a}+2{f^2}\dot{a}^2+{f^2}a\ddot{a}}{g^2}\delta_{ij}\\\nonumber
& &-\left(\frac{6{f^2}a\dot{a}\dot{g}+{a^2}f\dot{f}\dot{g}+{a^2}{f^2}\ddot{g}}{g^3}-\frac{4{a^2}{f^2}\dot{g}^2}{g^4}\right)\delta_{ij},\\
R&=&6f^2\left(\frac{\ddot{a}}{a}+\frac{\dot{a}\dot{f}}{af}+\frac{\dot{a}^2}{a^2}\right)\\\nonumber
&
&-6f^2\left(\frac{\ddot{g}}{g}+\frac{\dot{f}\dot{g}}{fg}+\frac{4\dot{a}\dot{g}}{ag}-\frac{3\dot{g}^2}{g^2}\right).
\end{eqnarray}
Finally, substituting all the terms above into Eq. (\ref{eq4}) we
obtain the modified FRW equations as
\begin{eqnarray}
(H-\frac{\dot{g}}{g})^2&=&\frac{8\pi{G}(\varepsilon)\rho}{3f^2}+\frac{\Lambda(\varepsilon)}{3f^2},\label{MFWR1}\\
\dot{H}+\frac{\dot{g}^2}{g^2}-\frac{\ddot{g}}{g}&=&-\frac{4\pi{G}(\varepsilon)(\rho+P)}{f^2}-(H-\frac{\dot{g}}{g})\frac{\dot{f}}{f},
\label{MFWR2}\end{eqnarray} where $H=\dot{a}/a$ is the Hubble
parameter and $\rho$, $P$ denote the energy density and the
pressure of perfect fluids respectively.

Furthermore, the conservation law for the energy-momentum tensor
reads as $\nabla(\varepsilon)^\mu{T_\mu{_\nu}}=0$, where the
covariant derivative $\nabla(\varepsilon)^{\mu}$ is energy
dependent as well. Plugging all the components of the connection
into this equation leads to a modified conservation equation as
\begin{equation}
\dot{\rho}+3(H-\frac{\dot{g}}{g})(\rho+P)=-\frac{\dot{\Lambda}(\varepsilon)}{8\pi{G}}-\frac{\dot{G}}{G}\rho.\label{MCE}
\end{equation}
Obviously the conservation equation (\ref{MCE}) is not independent
and can be derived from Eqs.(\ref{MFWR1}) and (\ref{MFWR2}).

We now turn back to Eq. (\ref{eq5}) and look at the term
$a(t)/g^{2}$, which depends on the energy of probes through the
factor $g^{-2}$. Obviously, at low energy limit $g^{2}\rightarrow
1$, it is just the ordinary scale factor measured by observers. The
cosmological singularity occurs in standard model when one tracks
back to the origin of the universe and insists to take this as the
physical scale factor, namely $a(t)\rightarrow 0$ as $t\rightarrow
0$( or at some finite time). However, in rainbow gravity formalism
we find that as the energy of those particles \textit{evolving along
with spacetime geometry} increase at the very early stage of the
universe, the factor $g^{2}$ may be far from the unit and its
effects can not be ignored. Then any possible measurement by
\textit{those} particles  is energy dependent such that the actual
and physical scale factor measured by a probe depends on its energy
$\varepsilon$ through the function $g^{2}$, which is quite different
from the ordinary scale factor $a(t)$. As a matter of fact, in this
formalism as $a(t)\rightarrow 0$, the rainbow function $g^{-2}$
probably becomes divergent such that the total effective scale
factor probed by particles may be finite, providing a mechanism of
avoiding the singularity. Therefore we define an effective scale
factor $a_{eff}=a/g$ and subsequently the effective Hubble parameter
as $H_{eff}=\dot{a}_{eff}/a_{eff}=H-\dot{g}/g$. They are the
physical parameters probed by those matter evolving with space time
geometry in very early universe.  Through this paper we will focus
on analyzing the solutions to the effective scale factor $a_{eff}$.

\section{The big bounce solutions with $\Lambda=0$}
Now we are going to look for big bounce solutions to the modified
$FRW$ equations (\ref{MFWR1}) and (\ref{MFWR2}). For simplicity
through this paper we will treat the Newton's constant $G$ to be
energy independent. Next we need to specify the rainbow functions
$f(\varepsilon)$ and $g(\varepsilon)$. It has been pointed out in
\cite{Ling06az} that the effective $FRW$ equations appearing in
loop quantum cosmology\cite{Ashtekar06rx,Ashtekar06uz,Singh06im}
can be heuristically derived from the framework of rainbow gravity
once the modified dispersion relation is properly assigned. For
instance, the effective equation for $K=0$ and $\Lambda=0$ in loop
quantum gravity has the form \f H^2={8\pi G\over 3}\rho \left(
1-{\rho\over \rho_c}\right)\label{lqc1}\ff where $\rho_c\sim
l_p^{-4}$. They may be obtained from the side of rainbow gravity
if we naively set that \f f^2=(1-l_p^4\varepsilon^4)^{-1},\ \ \ \
\ \ \ \ \ g^2=1,\label{ansatz}\ff and
$\varepsilon^4\sim\bar{\varepsilon}^4\sim \rho$. This
identification presumably indicates a relation between the rainbow
gravity formalism and the effective theory of loop quantum gravity
at the semi-classical level. As a result it appears that the big
bounce solutions should be obtained in rainbow universe
straightforwardly. However, as pointed out in\cite{Ling06az}, this
naive identification involves more subtleties to clarify from the
side of doubly special relativity. The reason is that the
modification of dispersion relations will provide corrections to
the expectation value of statistical quantities of an ensemble as
well. If a modified dispersion relation does not manifestly
provide an upper bound on either the momentum or energy of a
single particle, it would lead to a divergent density of states as
the energy approaches to the Planck scale, which can be seen from
the following definition \cite{Alexander01ck,Ling06az} \fa
g(\varepsilon)d\varepsilon &=& 2{V\over h^3}4\pi p^2dp \nonumber\\
&=& {8\pi V\over h^3c^3}f^3\left( 1+\varepsilon{f'\over f}
\right)\varepsilon^2 d\varepsilon, \label{dos}\ffa  where prime
denotes the derivative with respect to the energy $\varepsilon$.
To preserve the finiteness of the mean value of statistical
quantities with the use of Eq.(\ref{dos}), it seems that some
certain cutoff of the energy should be introduced by hand, which
obviously sounds not satisfying. A more detailed analysis shows
that this difficulty does arise when one intends to derive the big
bounce solution in rainbow universe through the ansatz
(\ref{ansatz}). To avoid this weak point, we propose the following
two sorts of modified dispersion relations and then discuss the
solutions to the corresponding modified $FRW$ equations.

\subsection{the case of $f^2=g^2=\frac{1}{1-l_p^4\varepsilon^4}$}

The first sort of rainbow functions we would like to adopt is
\begin{equation}
f^2=g^2=\frac{1}{1-l_p^4\varepsilon^4}\label{eq15}.
\end{equation}
Obviously in any case of $f^2=g^2$, the ordinary relation
$d\varepsilon=dp$ is preserved for massless particles such that
all the statistical quantities will not receive corrections due to
the modification of dispersion relations. In particular, the
equation of state $P=\omega\rho$ and the relation $\rho\sim
\bar{\varepsilon}^4$ still hold for those particles.

Next we identify the energy appearing in rainbow metric with the
statistical mean value of all massless radiation particles, namely
$\varepsilon\sim\bar{\varepsilon}$. This is because we are
concerned with the ``average'' effect of radiation particles which
dominate the very early universe, rather than picking up any
specific particle from the radiation at random. The strategy of
treating all radiation particles as an ensemble and considering
their statistical effects has previously been applied to
investigate the thermodynamics of black holes as well
\cite{Adler01vs,Ling05bp,Galan06by}. As a result, the rainbow
functions of $f$ and $g$ can be finally expressed as
\begin{equation}
f^2=g^2=\frac{1}{1-l_p^4\bar{\varepsilon}^4}=\frac{1}{1-l_p^4\rho}.
\end{equation}
Correspondingly the modified FRW equations read as
\begin{eqnarray}
\frac{\dot{\rho}}{\rho}&=&-3H_{eff}(1+\omega)\label{eq17},\\
H_{eff}^2&=&\frac{8\pi{G}\rho}{3}(1-l_p^4\rho)\label{eq18},
\end{eqnarray}
which are exactly the effective $FRW$ equations arising in loop
quantum gravity. Without surprise when $\omega$ is a constant but
not equal to $-1$, we have the big bounce solution as
\begin{equation}
\rho=\frac{1}{(Ct)^2+l_p^4}\label{eq24},
\end{equation}
and
\begin{equation}
a_{eff}=[(Ct)^2+l_p^4]^\frac{1}{3(1+\omega)},\label{eq21}
\end{equation}
where $C=\frac{3(1+\omega)}{2}\sqrt{\frac{8\pi{G}}{3}}$. It is
easy to find that the energy density $\rho$ is bounded at the big
bounce of the universe, namely
\begin{eqnarray}
\rho\rightarrow\frac{1}{l_p{^4}},\quad\mbox{as}\quad
t\rightarrow0.\label{eq25}
\end{eqnarray}
Therefore, in rainbow universe the cosmological singularity can be
avoided. For explicitness, we plot the evolution of the effective
scale factor for two special cases with $\omega=-2/3$ and
$\omega=1/3$, as illustrated in Fig.1 and Fig.2 respectively.

\begin{figure}
\center{
\includegraphics[scale=0.75]{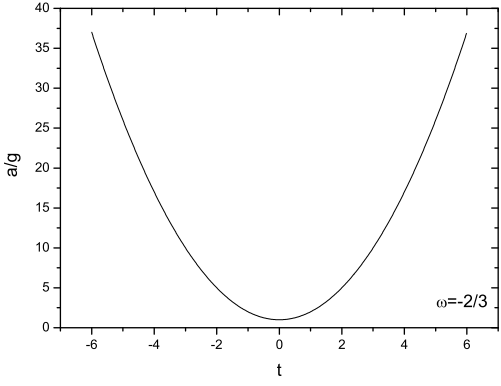}
\caption{The effective scale factor $a_{eff}$ vary with the
cosmological time when $f^2=g^2=\frac{1}{1-l_p^4\rho}$,
$\omega=-\frac{2}{3}$}}
\end{figure}

\begin{figure}
\center{
\includegraphics[scale=0.75]{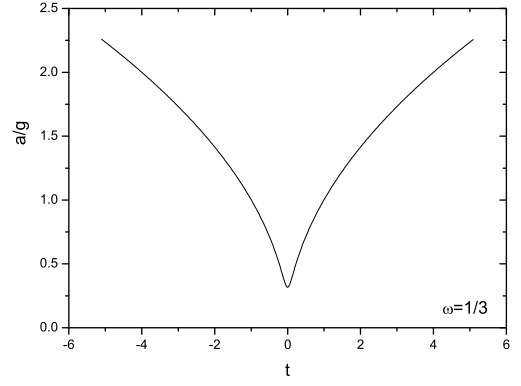}
\caption{The effective scale factor $a_{eff}$ vary with the
cosmological time when $f^2=g^2=\frac{1}{1-l_p^4\rho}$,
$\omega=\frac{1}{3}$}}
\end{figure}

From Eq.(\ref{eq21}), we find there is always a big bounce for the
very early universe if $\omega>-1$. However,  some issues arise
when one attempts to understand the behavior of the unverse at the
moment $t=0$. Though both the energy density and the effective
scale factor are finite at $t=0$, the rainbow functions
$f^2=g^2=\frac{1}{1-l_p{^4}\rho}$ approach to infinity as
$\rho\rightarrow 1/l_p^4$ such that the time-like component of the
rainbow metric vanishes, implying the degeneracy of the metric at
that time. Such difficulty is general for this sort of modified
dispersion relations and can not be avoided by inserting any
adjustable parameter like $f^2=g^2=\frac{1}{1-\eta l_p{^4}\rho}$.
To overcome this unsatisfactory point, we consider another sort of
modified dispersion relations in next subsection.

\subsection{the case of $f=g=1+l_p^4\bar{\varepsilon}^4$}

Now we attempt to take the rainbow functions $f$ and $g$ as
\begin{equation}
f=g=1+l_p^4\bar{\varepsilon}^4.\label{eq27}
\end{equation}
Then the effective FRW equation reads as
\begin{equation}
H_{eff}^2=\frac{8\pi{G}}{3}\frac{\rho}{(1+l_p^4\rho)^2}.\label{eq28}
\end{equation}
Combining this equation and the conservation equation (\ref{MCE})
leads to the evolution equation of the energy density
\begin{equation}
-\frac{\dot{\rho}}{3(1+\omega)\rho}=\sqrt{\frac{8\pi{G}}{3}}\frac{\sqrt{\rho}}{1+l_p{^4}\rho}.\label{eq29}
\end{equation}
The solution to this equation is
\begin{equation}
\rho=\frac{(Ct)^2+2l_p{^4}-\sqrt{(Ct)^4+4l_p{^4}(Ct)^2}}{2l_p{^8}},\label{eq31}
\end{equation}
where $C$ is a constant. Again the energy density is bounded and
approaches to $\frac{1}{l_p^4}$ as $t\rightarrow0$. Especially, in
this case both rainbow functions $f$ and $g$ at $t=0$ are finite
such that the rainbow metric is well defined even at the moment of
passing through the big bounce. Furthermore, we may obtain the
effective scale factor as
\begin{equation}
a_{eff}=\left[\frac{(Ct)^2+2l_p{^4}-\sqrt{(Ct)^4+4l_p{^4}(Ct)^2}}{2l_p{^8}}\right]^{-\frac{1}{3(1+\omega)}}.\label{eq32}
\end{equation}
It is also bounded and approaches to $l_p^{\frac{8}{3(1+\omega)}}$
as $t\rightarrow0$, thus a big bounce occurs whenever $\omega\neq
-1$. For instance, we illustrate its evolution with $\omega=-2/3$
in Fig.3. From this figure, however, we see that the evolution at
the moment of big bounce is {\it not} smooth. This comes from the
fact that the solution (\ref{eq32}) contains a square root such
that $\dot{a}_{eff}$ is not continuous at $t=0$.

\begin{figure}
\center{
\includegraphics[scale=0.75]{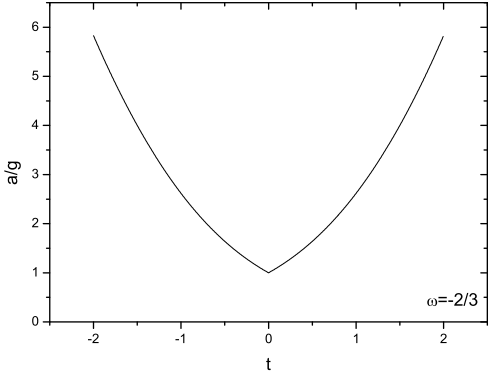}
\caption{The effective scale factor $a_{eff}$ vary with the
cosmological time when $f=g=1+l_p^4\rho$, $\omega=-\frac{2}{3}$}}
\end{figure}
In next section, we intend to demonstrate that all the
unsatisfactory points appearing in this section can be overcome by
introducing a non-zero cosmological constant.

\section{The big bounce in the presence of the cosmological constant }

In the framework of rainbow gravity, it is reasonable to expect
that both the Newton's constant and the cosmological constant are
dependent on the energy of probes\cite{Magueijo02xx}. Here we only
consider the possible modification of the cosmological constant
$\Lambda(\varepsilon)$. Introduce a third rainbow function $h$
such that $\Lambda(\varepsilon)=h^2(\varepsilon)\Lambda$ where
$\Lambda$ is a bare cosmological constant, the modified FRW
equation (\ref{MFWR1}) can be written as
\begin{equation}
H_{eff}^2=\frac{8\pi{G}\rho}{3f^2}+\frac{h^2(\varepsilon)\Lambda}{3f^2}.\label{eq33}
\end{equation}

To obtain the big bounce solutions and compare them with the
results in previous section, we fix the functions $f^2$ and $g^2$
as $f^2=g^2=\frac{1}{1-l_p^{4}\rho}$ and assume
$h^2=1+\lambda\rho$. In this setting Eq.(\ref{eq33}) becomes
\begin{equation}
H_{eff}^2=\frac{8\pi{G}}{3}\left[\rho(1-l_p^{4}\rho)+(1+\lambda\rho)(1-l_p^{4}\rho)\frac{\Lambda}{8\pi{G}}\right].\label{eq34}
\end{equation}
Combining Eqs. (\ref{MCE}) and (\ref{eq34}), we have
\begin{equation}
\frac{\dot{\rho}}{\rho\sqrt{-(l_p^{4}+\frac{\lambda\Lambda{l_p^{4}}}{8\pi{G}})\rho^2+[1+\frac{\Lambda}{8\pi{G}}(\lambda-l_p^{4})]\rho+\frac{\Lambda}{8\pi{G}}}}\equiv-M,\label{eq35}
\end{equation}
where $M\equiv
3(1+\omega)\sqrt{\frac{8\pi{G}}{3}}\frac{8\pi{G}}{8\pi{G}+\lambda\Lambda}$.
Here, we require $1+\frac{\lambda\Lambda}{8\pi{G}}\neq0$.
Introducing new variables
\begin{equation}
x\equiv\frac{1}{\rho},\ \ \ \ \
\rho_{\Lambda}\equiv\frac{\Lambda}{8\pi{G}}.\label{eq36}
\end{equation}
Eq.(\ref{eq35}) can be changed into
\begin{equation}
\frac{\dot{x}}{\sqrt{\rho_{\Lambda}x^2+[1+(\lambda-l_p^{4})\rho_{\Lambda}]x-l_p^{4}(1+\lambda\rho_{\Lambda})}}=M.\label{eq37}
\end{equation}
Next we present solutions to this equation in accord with the sign
of the cosmological constant $\Lambda$. Here for convenience, we
define T as
\begin{equation}
T=\rho_{\Lambda}x^2+[1+(\lambda-l_p^{4})\rho_{\Lambda}]x-l_p^{4}(1+\lambda\rho_\Lambda).
\end{equation}

\subsection{The case of $\Lambda>0$}

When $\Lambda>0$, we obtain the solution to Eq. (\ref{eq37}) as
\begin{eqnarray}
 M\sqrt{\rho_{\Lambda}}t&=&\ln [
2\rho_{\Lambda}x+[1+(\lambda-l_p^{4})\rho_{\Lambda}]+2\sqrt{\rho_{\Lambda}}\nonumber\\
&&\sqrt{T}]-\ln{N}, \label{38}\end{eqnarray}
 where to make the big bounce occur at $t=0$ we have
introduced an integral constant $N>0$ which is defined as
\begin{equation}
N^2=[1+(\lambda-l_p^{4})\rho_{\Lambda}]^2+4l_p^{4}\rho_{\Lambda}(1+\lambda\rho_{\Lambda})>0.\label{eq39}
\end{equation}
If we consider $\lambda>-\frac{1}{\rho_{\Lambda}}$, Eq.
(\ref{eq39}) has been satisfied. It is very useful to rewrite the
solution in Eq.(\ref{38}) as
\begin{eqnarray}
x&=&\frac{1}{4\rho_{\Lambda}}\exp{(M\sqrt{\rho_{\Lambda}}t+\ln{N})}\nonumber\\
&+&\frac{1}{4\rho_{\Lambda}}N^2\exp{[-(M\sqrt{\rho_{\Lambda}}t+\ln{N})]}\nonumber\\
&-&[1+(\lambda-l_p^{4})\rho_{\Lambda}]\frac{1}{2\rho_{\Lambda}}.\label{eq40}
\end{eqnarray}
It is easy to show that $x$ has the minimal value at $t=0$,
\begin{eqnarray}
x_{min}&=&\frac{1}{2\rho_{\Lambda}}[N-1-(\lambda-l_p^{4})\rho_{\Lambda}]\nonumber\\
&\approx&\frac{l_p^{4}(1+\lambda\rho_{\Lambda})}{1+(\lambda-l_p^{4})\rho_{\Lambda}}.\label{eq41}
\end{eqnarray}
If the parameter $\lambda$ satisfies the condition
\begin{equation}
\lambda>-\frac{1}{\rho_{\Lambda}}+l_p^{4},\label{eq43}
\end{equation}
then we find the energy density $\rho<\frac{1}{l_p^{4}}$ at the big
bounce. As a matter of fact, in rainbow gravity formalism it is
reasonable to expect the parameter to be an order of $|\lambda|\sim
l_p^4$, the condition (\ref{eq43}) can be easily satisfied.
Therefore, in the presence of a cosmological constant the rainbow
function $f$ will never diverge such that the rainbow metric is
always well defined even at the big bounce. Moreover, the effective
scale factor evolves as
\begin{eqnarray}
a_{eff}&=&\left\{\frac{1}{4\rho_{\Lambda}}\exp{(M\sqrt{\rho_{\Lambda}}t+\ln{N})}\right.\nonumber\\
&+&\left.\frac{1}{4\rho_{\Lambda}}N^2\exp{[-(M\sqrt{\rho_{\Lambda}}t+\ln{N})]}\right.\nonumber\\
&-&\left.[1+(\lambda-l_p^{4})\rho_{\Lambda}]\frac{1}{2\rho_{\Lambda}}\right\}^{\frac{1+\lambda\rho_{\Lambda}}{3(1+\omega)}}.\label{eq44}
\end{eqnarray}
We plot its evolution in Fig.4, from which we find the big bounce
occurs at $t=0$.

\begin{figure}
\center{
\includegraphics[scale=0.75]{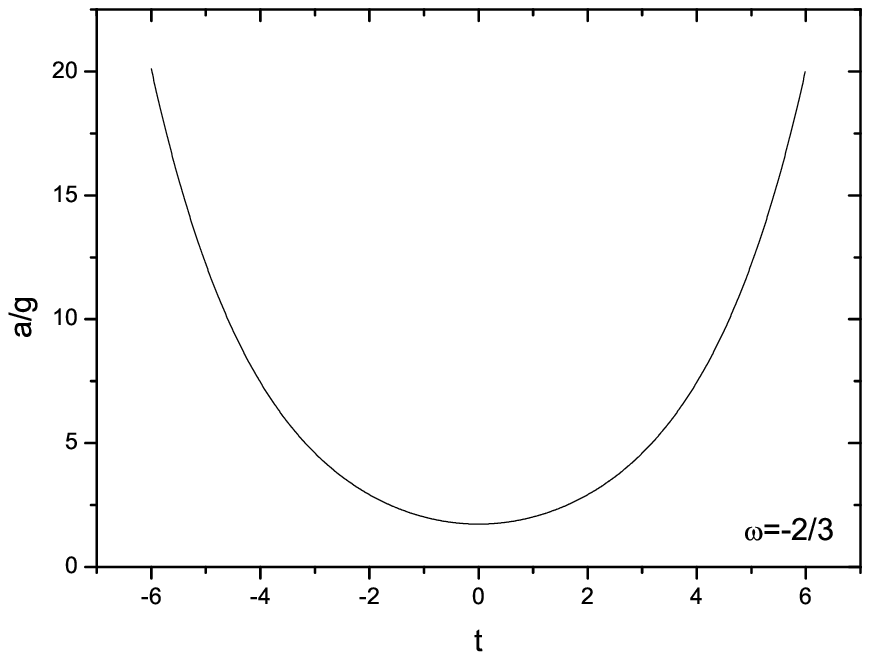}
\caption{The effective scale factor $a_{eff}$ varies with the
cosmological time when $\Lambda>0$.}}
\end{figure}

\subsection{The case of $\Lambda<0$}
When $\rho_{\Lambda}<0$, we obtain the solution by integrating
Eq.(\ref{eq37}),
\begin{eqnarray}
Mt&=&\sqrt{\frac{-1}{\rho_{\Lambda}}}
\arcsin{\left\{\frac{-2\rho_{\Lambda}x-[1+(\lambda-l_p^{4})\rho_{\Lambda}]}{N}\right\}}\nonumber\\
&+&\frac{\pi}{2}\sqrt{\frac{-1}{\rho_{\Lambda}}},\label{eq47}
\end{eqnarray}
where $\frac{\pi}{2}\sqrt{\frac{-1}{\rho_{\Lambda}}}$ is
introduced as an integral constant such that the big bounce shifts
to $t=0$. Similarly, we rewrite this solution as
\begin{equation}
x=-\frac{1}{2\rho_{\Lambda}}\left\{N\sin{(M\sqrt{-\rho_{\Lambda}}t-\frac{\pi}{2})}+[1+(\lambda-l_p^{4})\rho_{\Lambda}]\right\}.\label{eq48}
\end{equation}
Obviously $x$ has the minimal value at $t=0$,
\begin{eqnarray}
x_{min}&=&\frac{1}{2\rho_{\Lambda}}[N-1-(\lambda-l_p^{4})\rho_{\Lambda}]\nonumber\\
&\approx&\frac{l_p^{4}(1+\lambda\rho_{\Lambda})}{1+(\lambda-l_p^{4})\rho_{\Lambda}}.\label{eq41}
\end{eqnarray}
If the parameter $\lambda$ is constrained as
\begin{equation}
-\frac{1}{\rho_{\Lambda}}<\lambda<-\frac{1}{\rho_{\Lambda}}+l_p^{4},\label{eq50}
\end{equation}
then it is guaranteed that the energy density will never exceed the
Planck energy density, namely $\rho<l_p^{-4}$ and the rainbow metric
is always well defined.

The solution of the effective scale factor is
\begin{eqnarray}
a_{eff}&=&\{-\frac{1}{2\rho_{\Lambda}}[N\sin{(M\sqrt{-\rho_{\Lambda}}t
-\frac{\pi}{2})}\nonumber\\&+&(1+(\lambda-l_p^{4})\rho_{\Lambda})]\}^{\frac{1+\lambda\rho_{\Lambda}}{3(1+\omega)}}.\label{eq52}
\end{eqnarray}
We illustrate its evolution with the cosmological time in Fig.5.
This is an oscillation solution containing many big bounces.

\begin{figure}
\center{
\includegraphics[scale=0.75]{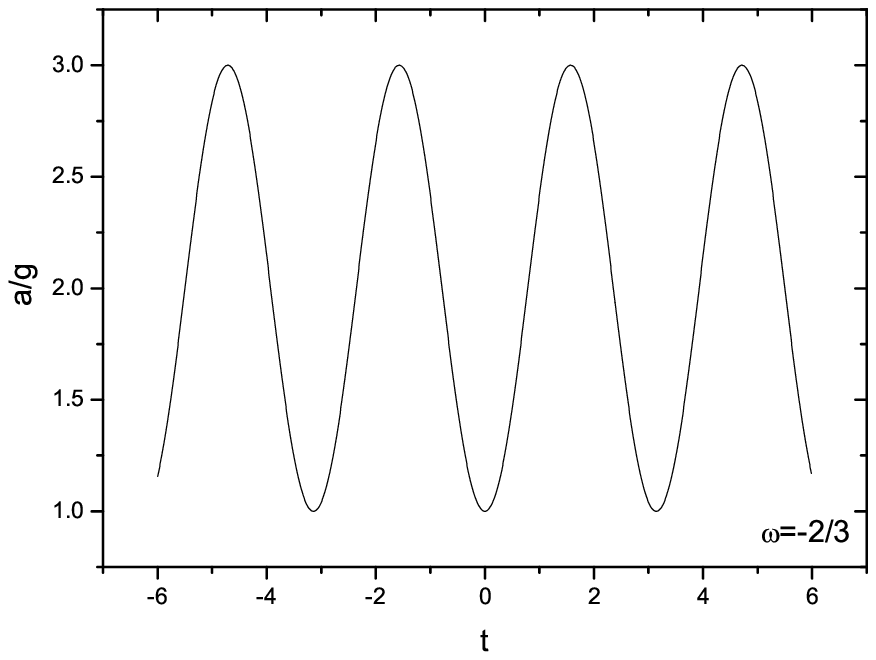}
\caption{The effective scale factor $a_{eff}$ varies with the
cosmological time when $\Lambda<0$.}}
\end{figure}

Similar big bounce solutions can be found in the context of  brane
scenarios and effective loop quantum
cosmology\cite{Stachowiak06uh,Mielczarek08zv}, where they
considered a $\rho^2$ -type modification to the Friedmann equation
due to the brane effects or discreteness of quantum geometry. Here
we obtain the big bounce solutions by considering the quantum
gravity effects of probes. By appropriately choosing the rainbow
functions $f^2$, $g^2$ and $h^2$ we find that $\rho^2$-type
modification to the Friedmann equation can be achieved in the
framework of rainbow universe as well. But in contrast to brane
cosmology or loop quantum gravity where the modification emerges
from the brane effects or quantum geometry such that the
coefficients in all correction terms are uniquely fixed, here we
treat the rainbow functions $f^2$, $g^2$ and $h^2$ as independent
modification quantities. It is also interesting to notice that in
loop quantum gravity the lattice regularization need to be refined
in order to assure the big bounce occurring at the Planck
scale\cite{Ashtekar06rx,Ashtekar06uz,Singh06im}. While in rainbow
gravity formalism, we find that the magnitude of the effective
scale factor at the big bounce is adjusted by the parameter
$\lambda$ in rainbow function $h$ which is assumed to be the order
of $l_p^4$. We may also compare our solutions with those appearing
in low energy effective theory of string theory, for instance
\cite{Fabris02pm}, where the existence of no-singular solutions is
quite relevant to the dilatonic coupling constant. In our cases
the specific form of rainbow functions plays a similar role, but
the presence of the cosmological constant makes it possible to
obtain bouncing solutions as well as oscillatory solutions.

\section{Conclusions}
In this paper we have constructed a phenomenological model for the
very early universe in the framework of rainbow gravity, in which
the quantum gravity effects of probes on spacetime background is
taken into account at the semi-classical level. Starting from the
rainbow Einstein's equation, we derived the modified $FRW$
equations by considering the cosmological evolution of particles.
Furthermore, given appropriate rainbow functions we find that the
big bounce solutions to the modified $FRW$ equations can be
obtained. Notably, to obtain a well defined rainbow metric at the
moment of the big bounce, we find it seems necessary to introduce
a cosmological constant which depends on the energy of probes as
well. This is a quite interesting result. It implies that a
universe with a positive cosmological constant more likely
undergoes a big bounce at least at this phenomenological level. A
complete understanding on the behavior of the universe at the big
bounce probably calls for a complete quantum theory of gravity.
However, our semi-classical analysis presented here has shed light
on this issue and indicates that such quantum gravity effects can
contribute important corrections to the classical general
relativity so as to provide a more reasonable picture on the
origin of the universe. The validity of semi-classical analysis
has also been strengthened by recent progress in loop quantum
gravity in which the exact solutions to quantum equations are
remarkably well approximated by a naive extrapolation of some
semi-classical effective
equations\cite{Ashtekar06rx,Ashtekar06uz,Singh06im}.

Recently bouncing cosmology has received a lot of attentions as the
astronomical observation has released the cosmological data with
much higher precisions, which provide strong evidences for a nearly
scale-invariant spectrum of primordial density
fluctuations\cite{Komatsu08hk}. First of all, theoretical
investigation indicates that it is possible to generate such a
spectrum during the contracting phase in various bouncing models,
for instance see
references\cite{Tolley03nx,Peter02cn,Creminelli07aq}. Secondly, in
contrast to the slow-roll inflationary scenario in which the
non-Gaussian contribution to the density fluctuation spectrum is
greatly suppressed, some bouncing models require that the universe
experience a contracting phase before the hot Big Bang expansion
such that a large non-Gaussianity can be
predicted\cite{Biswas06bs,Lehners07wc,Lehners08my,Mielczarek08pf}.
The forthcoming observations maybe distinguish and rule out various
models of the origin of the universe. The non-singular bounce
solutions that we have obtained in this paper make it possible to
investigate the perturbation theory of the bouncing cosmology in
rainbow gravity formalism. Our investigation along this direction is
under progress.

Through the paper we only investigate some special cases in
rainbow gravity with specified functions of $f(\epsilon)$,
$g(\epsilon)$ and $h(\epsilon)$. However, we would like to point
out that a large class of rainbow functions may lead to bouncing
solutions if they could provide a $\rho^m$-type modification to
the Friedmann equation where $m$ is not necessarily fixed to be
square as in references. Such contributions would suppress the
effective Hubble parameter until it vanishes at the big bounce. Of
course, we expect the further tests relevant to Lorentz symmetry
would tell us  which is the proper one  among all the possible
modified dispersion relations.

It is completely possible to extend the scheme presented here to
more general cases, for instance when the  spatial metric is
non-flat and the Newton's constant is also energy dependent.

\begin{acknowledgments}
We would like to thank the members of the Center for Relativistic
Astrophysics and High Energy Physics at Nanchang University for
their lots of discussions and helpful suggestions.  This work is
partly supported by NSFC(Nos.10663001, 10875057), JiangXi SF(Nos.
0612036, 0612038), the key project of Chinese Ministry of
Education (No.208072) and Fok Ying Tung Eduaction
Foundation(No.111008). We also acknowledge the support by the
Program for Innovative Research Team of Nanchang University and
Jiangxi Young Scientists(JingGang Star) Program.
\end{acknowledgments}

%\newpage Just because of unusual number of tables stacked at end
\bibliography{apssamp}% Produces the bibliography via BibTeX.

\end{document}